\documentclass[pre, 11pt, preprint]{revtex4}

\usepackage{graphics}

\begin{document}

\title{On the nonlocal viscosity kernel of mixtures}

\author{Ben Smith}
\affiliation{Mathematics Discipline, Faculty of Engineering and
  Industrial Sciences, Swinburne University, 
  PO Box 218, Hawthorn, Victoria 3122, Australia.}
\author{J. S. Hansen}
\email{jschmidt@ruc.dk (Corresponding author)}
\affiliation{DNRF Centre ``Glass and Time'', IMFUFA, 
  Department of Sciences, Systems and Models, Roskilde University, 
  PO box 260, DK-4000 Roskilde, 
  Denmark} 
\author{B. D. Todd}
\email{btodd@swin.edu.au}
\affiliation{  
  Mathematics Discipline, Faculty of Engineering and
  Industrial Sciences, Swinburne University, 
  PO Box 218, Hawthorn, Victoria 3122, Australia.
}

\begin{abstract}
In this report we investigate the multiscale hydrodynamical response 
of a liquid as a function of mixture composition. This is done via a
series of molecular dynamics simulations where the wave vector
dependent viscosity kernel is computed for three mixtures each with
7-15 different compositions.  We observe that the nonlocal viscosity kernel is 
dependent on composition for simple atomic mixtures for all the wave
vectors studied here, however, for a model polymer melt
mixture the kernel is independent of composition for large wave
vectors. The deviation from ideal mixing is also studied. Here it is
shown that a Lennard-Jones mixture follows the ideal mixing rule
surprisingly well for a large range of wave vectors, whereas for both the
Kob-Andersen mixture and the polymer melt large deviations are
found. Furthermore, for the polymer melt the deviation is wave vector
dependent such that there exists a critical length scale at which the ideal
mixing goes from under-estimating to over-estimating the viscosity.
\end{abstract}


\maketitle

Hydrodynamics on very small length scales has become an 
important research area because it is believed to hold
the key to understand the many new phenomena observed in
nanofluidic devices. Recent studies~\cite{todd_2008, todd_2008_2}
have shown that the spatial correlations in the fluid reduce  
the shear stress compared to the stress predicted via a local
response function. The nonlocal response is described via
generalized hydrodynamics~\cite{alley_1983}, in which the response
function is a wave vector dependent quantity. 
The wave vector dependent viscosity, i.e. the viscosity kernel,
accounts for the momentum flux due to nonzero strain rate. It 
has been found for one component fluids through molecular dynamics
simulations and it was found that it follows a simple functional
form reasonably well for atomic, diatomic and polymer
fluids~\cite{alley_1983, hansen_2007_2, puscasu_2010_1,puscasu_2010_2}. By now
the effect of the multiscale response is understood fairly well for a
range of simple single component fluids~\cite{alley_1983,
  hansen_2007_2, puscasu_2010_1,puscasu_2010_2} and
glasses~\cite{furukawa_2009, puscasu_2010_3}.    

In order to investigate the multiscale hydrodynamical response as a
function of composition, we here evaluate the viscosity kernel for
three different two component mixtures using molecular dynamics. 
In details, these are (i) a Kob-Andersen (KA) mixture~\cite{kob_1994},
(ii) a Lennard-Jones (LJ) mixture using the 
Lorentz-Berthelot mixing rule, and (iii) a model polymer
melt mixture~\cite{kremer_1990}. In all the simulations the particles
interact through the Lennard-Jones cut and shifted potential 
$U_{LJ}(r_{ij}) = 4\epsilon \left[\left(\sigma/r_{ij}\right)^{12} -
  \left(\sigma/r_{ij}\right)^{6}\right] - U(r_c)$ for $r_{ij} \leq r_c$,  
where $r_{ij}$ is the distance between particle $i$ and $j$, 
$\sigma$ is a length scale, $\epsilon$ is an energy scale, 
$r_c$ is the interaction range (cut-off) and $U(r_c)$ is the
unshifted potential at $r_c$. The values of $\sigma$ and $\epsilon$ are
different for the KA and LJ mixtures depending on the pair of
particles that interact, see Table 
\ref{table:param}.   
\begin{center}
  \begin{table}
    \begin{tabular}{ccc} \\ \hline \hline 
      $\mbox{}$      & Kob-Andersen \hspace{0.5cm}& Lennard-Jones  \\ 
      \hline
      $\epsilon_{AA}/\epsilon_{AA}$ & 1            &       1        \\
      $\epsilon_{BB}/\epsilon_{AA}$ & 1/2          &       1/2      \\
      $\epsilon_{AB}/\epsilon_{AA}$ & 0.8          &   $\sqrt{1/2}$ \\
      $\sigma_{AA}/\sigma_{AA}$     & 1            &       1        \\  
      $\sigma_{BB}/\sigma_{AA}$     & 0.88         &       0.88     \\
      $\sigma_{AB}/\sigma_{AA}$     & 0.8          &       0.94     \\ 
      \hline \hline
    \end{tabular}
    \caption{\label{table:param}
      List of the Kob-Andersen and Lennard-Jones parameters used in
      this work.}
  \end{table}
\end{center}
The cut-off radius is set to $r_c = 2.5 \sigma$ for the KA and LJ
mixtures and $r_c = 2^{1/6} \sigma$ for the polymer melt. The latter is also
referred to as the Weeks-Chandler-Andersen pair potential~\cite{weeks_1971}.
In the polymer melt the particles (or beads) are bonded via
the Finite Extensible Nonlinear Elastic (FENE)
potential~\cite{kremer_1990} $U_{FENE} = -kR_0\ln\left[ 1 -
  \left(r_{ij}/R_0\right)^2\right]/2$,  
where $k=30 \epsilon/\sigma^2$ and $R_0=1.5 \sigma$. The polymer melt
is composed of two types of molecules, one with two beads, component B,
and one with ten beads, component A. 
In what follows we give all quantities in terms of Lennard-Jones
reduced units, for example, reduced distance $r_{ij}^*= r_{ij}/\sigma$
and number density $\rho^*= \rho\sigma^3$. For simplicity of notation,
we will hereafter omit the asterisk.

The simulations are carried out at an average reduced pressure of
$p=1$ and temperature $T=2.5$. The target pressure was obtained via an
anisotropic Berendsen barostat~\cite{berendsen_1984} such the simulation box
was extended in the $x$ direction only. The temperature was controlled
via a Nos\'{e}-Hoover thermostat~\cite{nose_1984,hoover_1985}. 

The expression for the wave vector dependent viscosity can be found 
from the generalized Navier-Stokes equation and is given in terms of
the transverse momentum current density autocorrelation function
$\widetilde{C}_\perp(k, t)$  
\cite{hansen_book_2006, palmer_1994}
\begin{eqnarray}
\eta(k, \omega) = 
\frac{\widetilde{C}_\perp(k, t=0) - i\omega
  \widehat{C}_\perp(k,\omega)}{\widehat{C}_\perp(k,\omega)k^2/\rho} \ , 
\end{eqnarray}
where $\rho$ is the mass density, $k$ is the $z$ component the wave
vector, i.e. 
\begin{eqnarray}
k=2\pi n/L_z\ , \ \ \ n=1,2,\ldots \label{eq:k} \ ,
\end{eqnarray}
where $L_z$ is the simulation box length in the $z$ direction, and
$\widetilde{C}_\perp(k, t) = \langle \widetilde{J}_y(k,0)
  \widetilde{J}_y(k,t)\rangle/V$.
The transverse momentum density is here defined via 
$\widetilde{J}_y(k,t) = \sum_{i=1}^{N} m_iv_y(t)e^{ikz},$
where $m_i$ and $v_y$ are the center of mass and center of mass
velocity of molecule or particle $i$. We note that
$\widehat{C}_\perp(k,\omega)$ is 
the Fourier-Laplace transform of $\widetilde{C}_\perp(k,t)$, that is, 
$\widehat{C}_\perp(k, \omega) = \int_0^\infty \widetilde{C}_\perp(k,t)
e^{i\omega t}dt$.
Also note, because the barostat is anisotropic and only varies the
simulation box in the $x$ direction $L_z$ is constant in
Eq.(\ref{eq:k}).  

\begin{center}
  \begin{figure}
    \scalebox{0.4}{
      \includegraphics{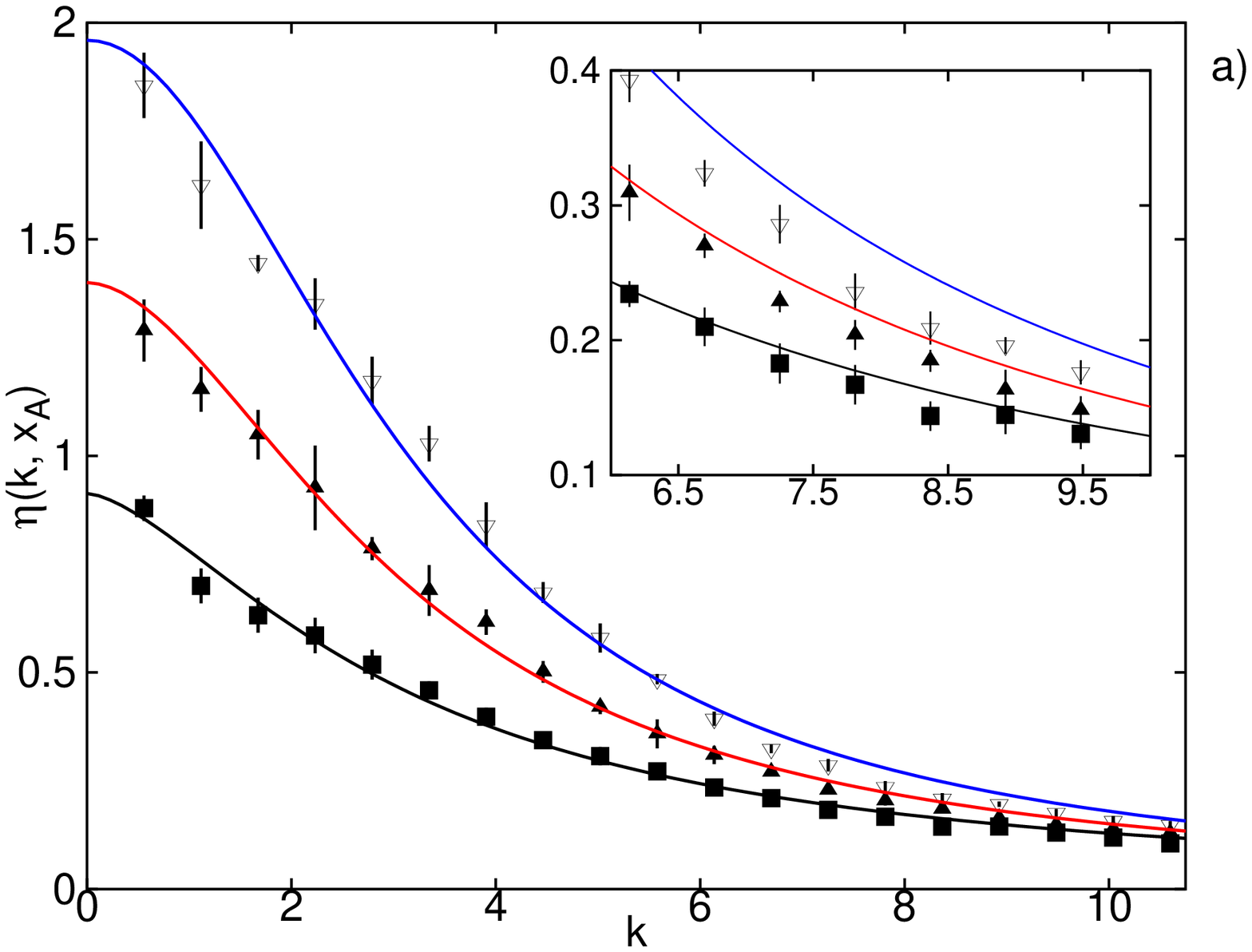}
    }
    \scalebox{0.4}{
      \includegraphics{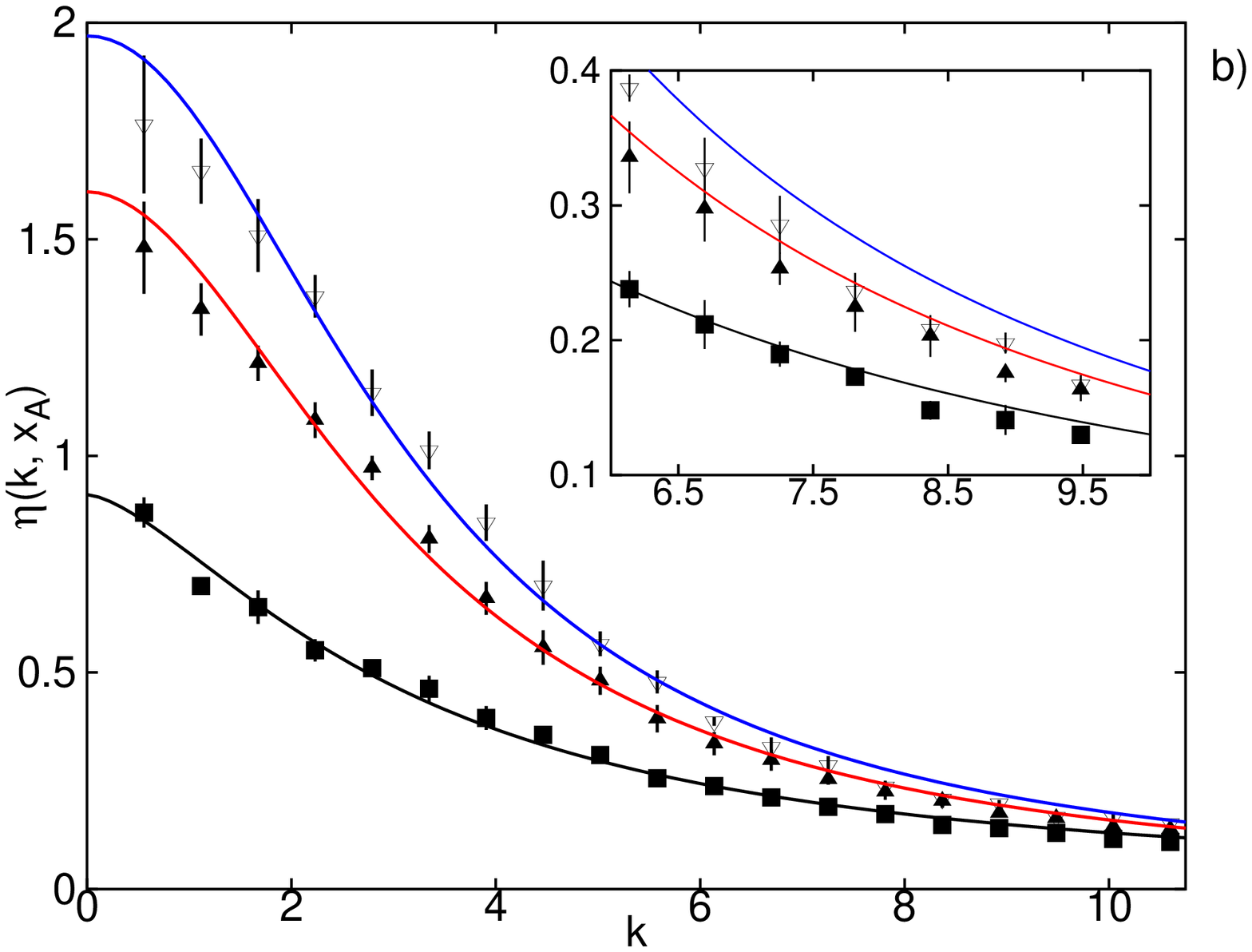}
    }
    \\
    \scalebox{0.4}{
      \includegraphics{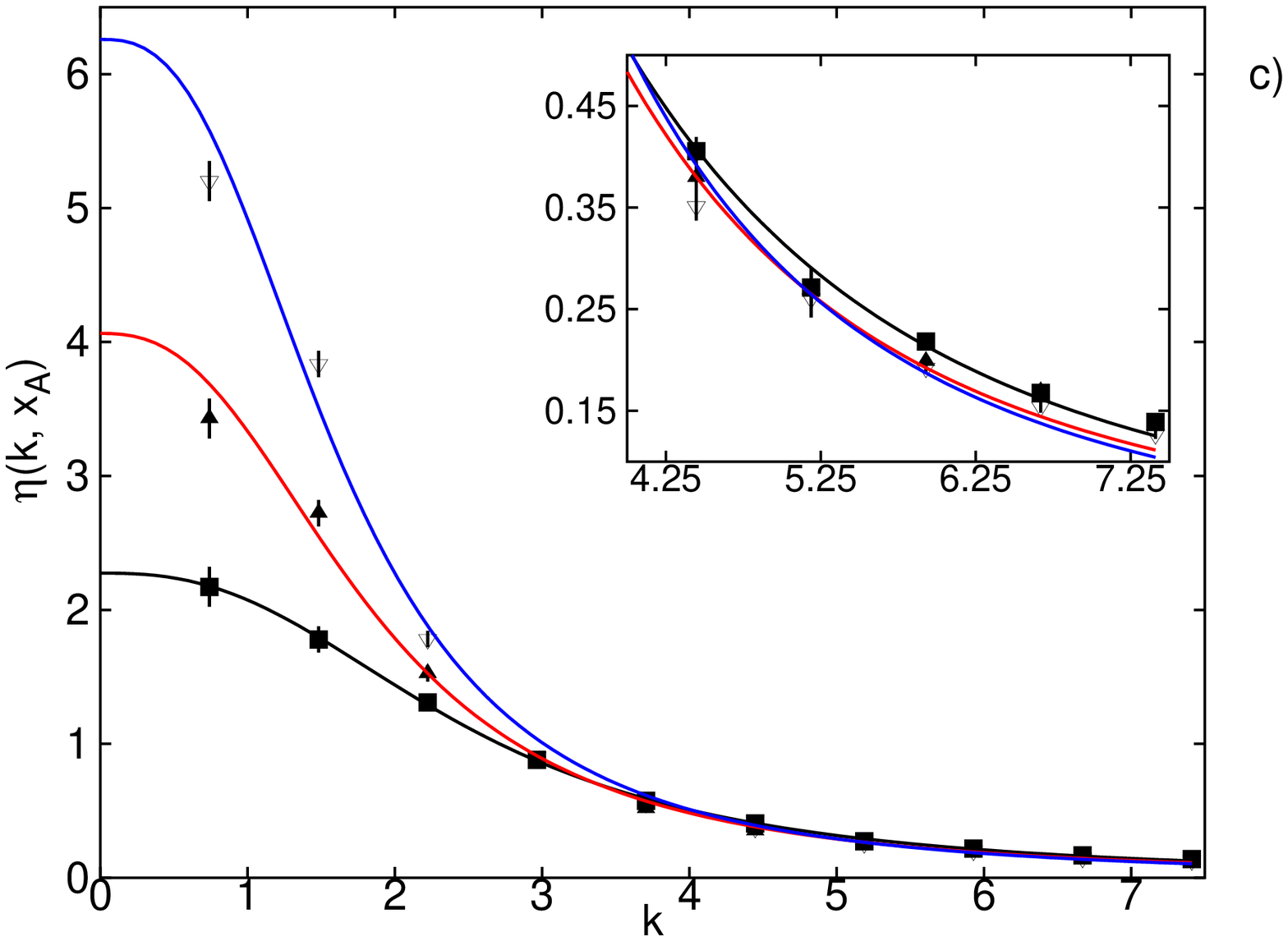}
    }
    \caption{
      \label{fig:1}
      Viscosity kernel data (symbols) for the Kob-Andersen mixture
      (a), the Lennard-Jones mixture (b) and the polymer mixture (c). 
      In (a) and (b): $x_A = 0$ (filled squares), 0.8 (upwards pointing
      triangles), and 1 (downwards pointing triangles). In (c): $x_A =
      0$ (filled squares), 0.25 (upwards pointing triangles) and 1
      (downwards pointing triangles). The error bars are the standard
      error. Lines represent the best fit of the data to
      Eq. (\ref{eq:lorentz}). The inset windows depict the kernels for
      large $k$.  
    }
  \end{figure}
\end{center}

In Fig. \ref{fig:1} we have plotted the viscosity kernel data in the
limit $\omega \rightarrow 0$ for different composition fractions of A, 
$x_A=N_A/N_{t}$ where $N_A$ is the number of A particles and $N_t$ is
the total number of particles. 
Recall, that in the polymer melt we
denote the 10 bead polymer A. From Fig. \ref{fig:1} it is observed
that for the KA and LJ mixtures the hydrodynamical response is dependent on
the exact composition for all length scales studied here.  This is not
the case for the polymer mixture: here the fluid response 
is largely independent of the fluid  composition for sufficiently
large wave vectors. In order to decrease the statistical error in
the further analysis we fit the data to a Lorentzian functional form  
\begin{eqnarray}
\label{eq:lorentz}
\eta(k, x_A) = \frac{\eta_0(x_A)}{1 + \alpha k^\beta} \ ,
\end{eqnarray}
where $\alpha$ and $\beta$ are fitting parameters that depend
on the fraction of A, $x_A$, and $\eta_0(x_A)$ is the zero frequency
viscosity for $k=0$. Eq. (\ref{eq:lorentz}) has been shown to
fit viscosity kernel data well for a range of different non-charged 
single component fluids~\cite{hansen_2007_2,puscasu_2010_1, puscasu_2010_2}. 
The result of the fitting is also depicted in Fig. \ref{fig:1} where    
the inset windows show the data and the fits for
large  values of $k$. It is seen that Eq. (\ref{eq:lorentz}) fits data
well for all three systems, however, we stress that the agreement is 
not satisfactory for large values of $k$ in the cases of
Kob-Andersen and Lennard-Jones mixtures. With this in mind we, will
from now on use the fitted values of the viscosity kernels
rather than the raw MD data. 

\begin{center}
  \begin{figure}
    \scalebox{0.4}{
      \includegraphics{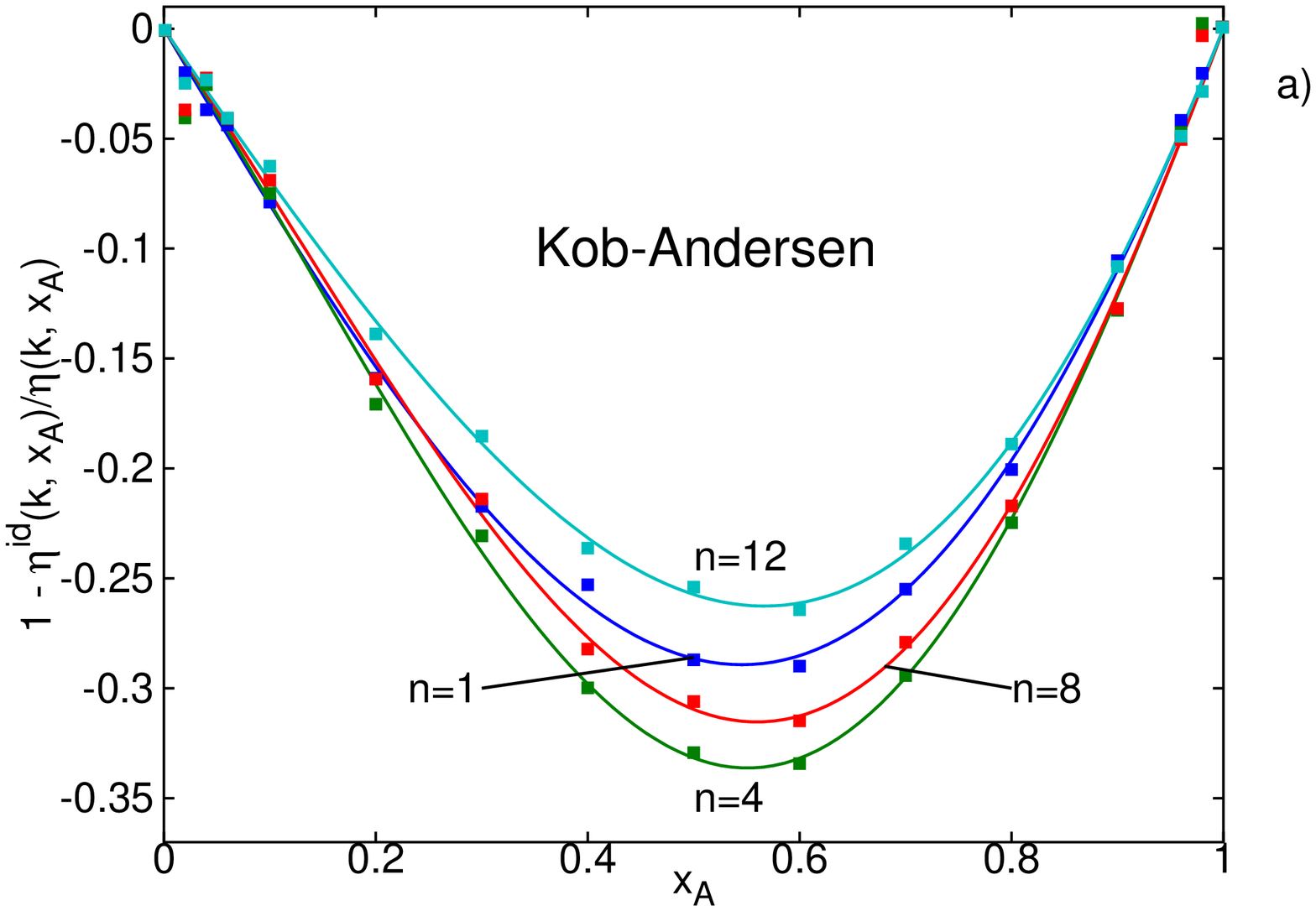}
    }
    \scalebox{0.4}{
      \includegraphics{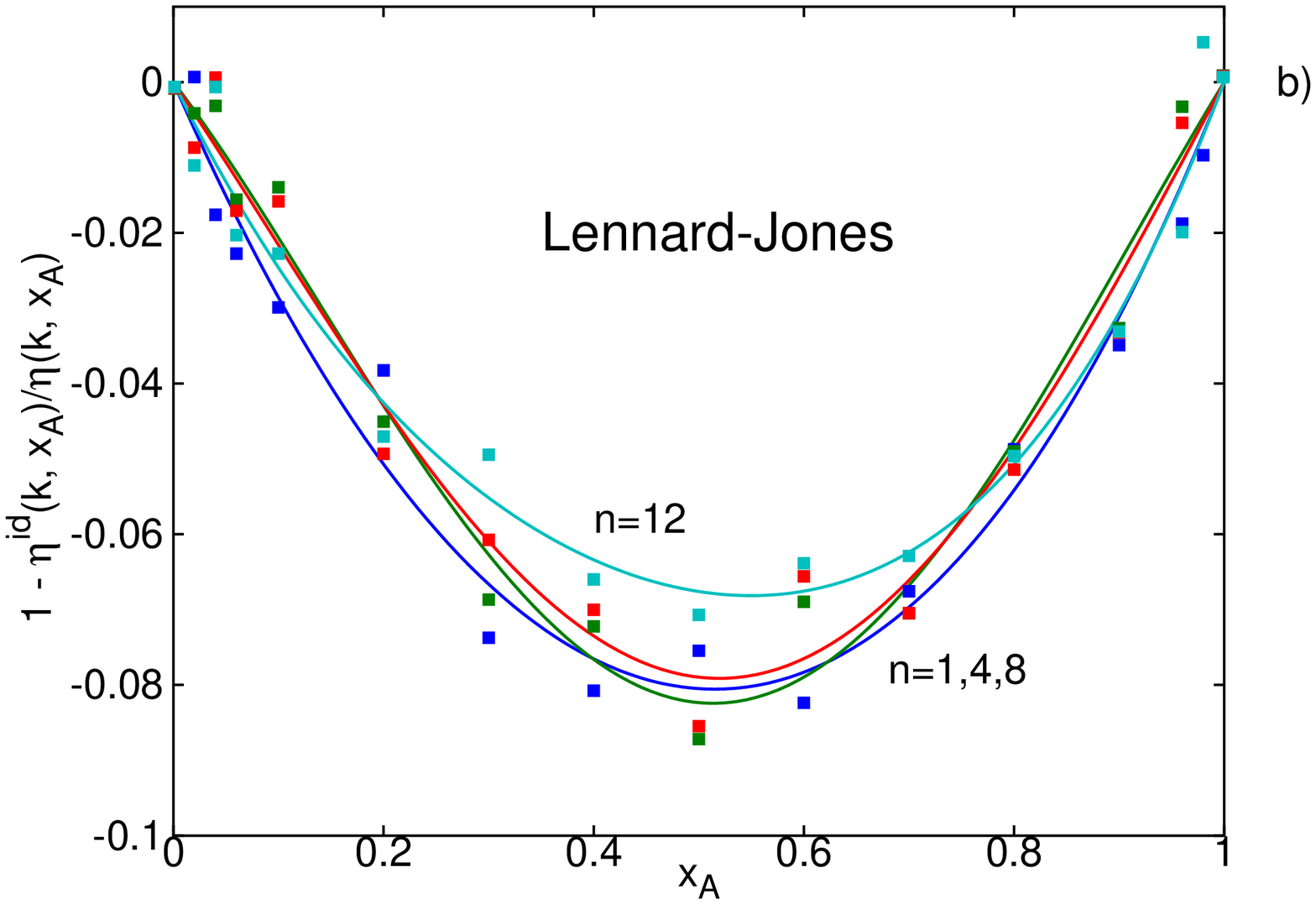}
    }
    \scalebox{0.4}{
      \includegraphics{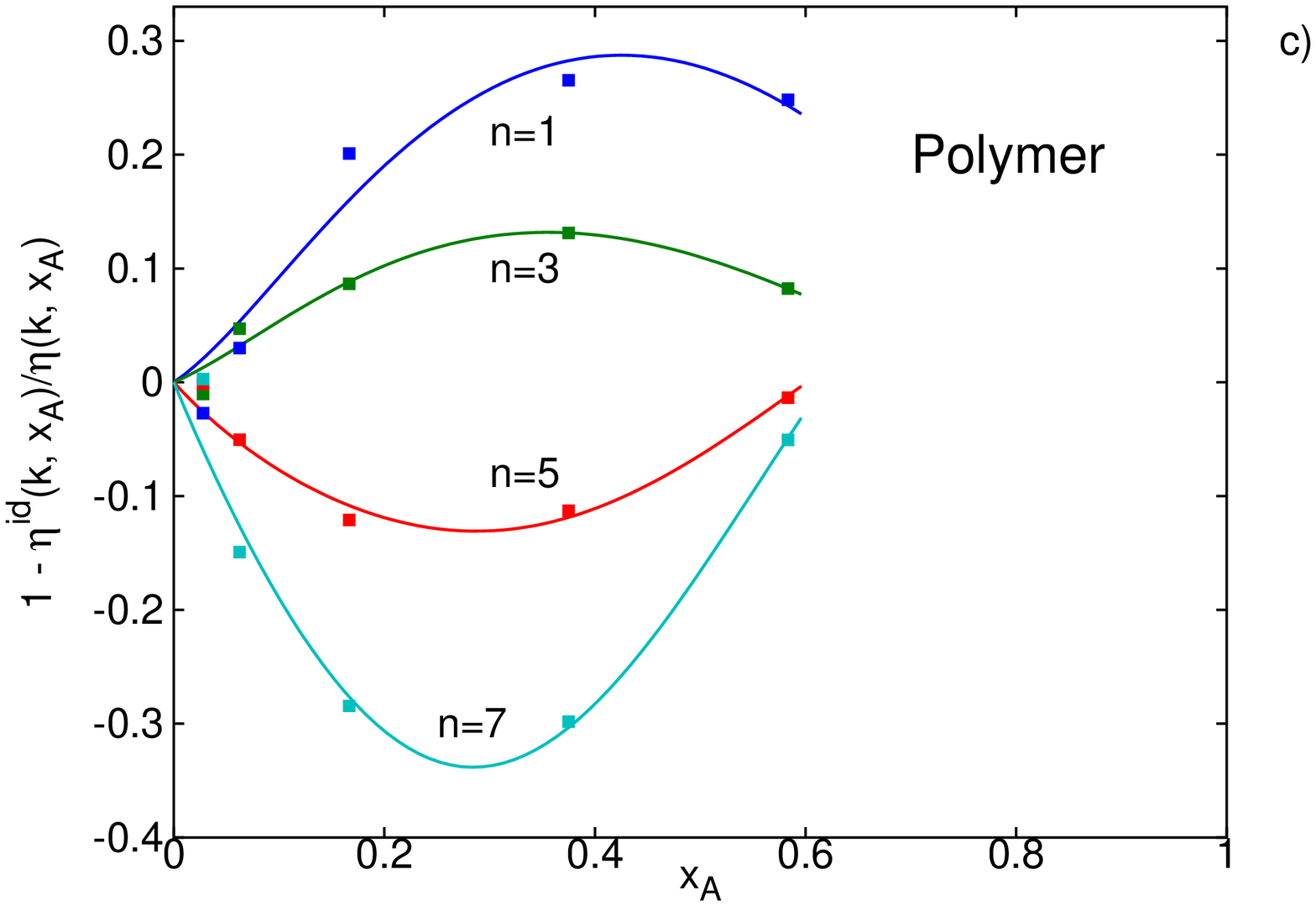}
    }
    \caption{
      \label{fig:3}Relative excess viscosity as function of
      composition and wave number, $n$. a) The Kob-Andersen mixture, b) the
      Lennard-Jones mixture and c) the polymer mixture.  Full li nes are
      the corresponding fits using Eqs.~(\ref{eq:mcallister}) and (\ref{eq:ideal}).    
    }
  \end{figure}
\end{center}

We can define the $k$ dependent excess viscosity as
\begin{eqnarray}
\eta^E(k, x_A) = \eta(k, x_A) - \eta^{\text{id}}(k, x_A) \ ,
\label{eq:excess}
\end{eqnarray}
where $ \eta^{\text{id}}(k, x_A)$ is the ideal part of the viscosity
kernel given by an Arrhenius type mixing rule~\cite{arrhenius_1887,kendall_1921}
\begin{eqnarray}
 \eta^{\text{id}}(k, x_A) = \eta_A(k)^{x_A} \eta_B(k)^{1-x_A} 
\label{eq:ideal}
\end{eqnarray}
Here $\eta_A(k)$ and $\eta_B(k)$ are the viscosity kernels of pure $A$
and $B$, respectively. The excess viscosity has been fitted to various
simple mixing models including the Kendall-Monroe
model~\cite{kendall_1921}, the Lederer model, see for example
Ref. \cite{lederer_1931},  and an extended version of the Grundberg-Nissan
model \cite{grunberg_1949}. We have found that the fourth order McAllister
model~\cite{mcallister_1960} fitted the data best. This model is originally 
written in terms of the dynamical viscosity $\nu=\eta/\rho$ 
and may readily be extended to include the wave vector dependency,
that is, 
\begin{eqnarray}
 \ln \nu(k, x_A) &=& x_A^4\ln[\nu_A(k)] +4x_A^3x_B\ln[M_{31}(k)]+ 
 6x_A^2x_B^2\ln[M_{22}(k)]+  
 \nonumber \\ &\mbox{}&
 4x_Ax_B^3\ln[M_{13}(k)] +x_B^4\ln[\nu_B(k)] - \ln(x_A +x_Bm_r)+ 
 \nonumber \\  &\mbox{}&
4x_A^3x_B\ln\left(\frac{3+m_r}{4}\right) + 
6x_A^2x_B^2\ln\left(\frac{1+m_r}{2}\right)+
 \nonumber \\  &\mbox{}& 
4x_Ax_B^3\ln\left(\frac{1+3m_r}{4}\right)+ 
x_B^4\ln(m_r) \ ,\label{eq:mcallister}
\end{eqnarray}
where $m_r$ is given by $m_B/m_A$ and $M_{31},M_{22}$ and $M_{13}$ are
the wave vector dependent 
McAllister coefficients. Recall, since we study binary mixtures
$x_B=1-x_A$. From Eq. (\ref{eq:mcallister}) one can easily extract the
McAllister excess kinematic viscosity using $M_{31},M_{22}$ and $M_{13}$ as
fitting parameters Rather than comparing the absolute excess
viscosities, we compare the relative deviation from ideal mixing using  
$\eta^E(k,x_A)/\eta(k,x_A) = 1 - \eta^{\text{id}}(k,x_A)/\eta(k,x_A)$.
This is done in Fig. \ref{fig:3}. 
We also note that the  normalized excess viscosity 
is a measure of the relative deviation from ideal mixing. 
Firstly, it is observed that the
KA and polymer mixtures feature large deviations from the
ideal mixing rule compared to the LJ mixture, that is, 
the Lorentz-Berthelot mixing rule features a more Arrhenius-like behavior
compared to the KA mixing rule for all wave lengths. Secondly,
for the KA and LJ mixtures the relative excess
viscosity varies very little with respect to wave vector 
meaning that the relative difference between the ideal mixing 
response function and the actual response is weakly wave
vector dependent. For the polymer mixture, Fig. \ref{fig:3} c), we
observe that the relative excess viscosity is strongly wave vector dependent. 
This is also indicated in Fig. \ref{fig:1} c) where it was observed
that the kernel is independent of composition for large $k$,
but not for small $k$. This complex wave vector dependency means that
there exists a critical length scale $l_c$ at which the ideal mixing
term goes from under-estimating to over-estimating the
viscosity. For the melt studied here it happens around
$l_c = 1.7$. This behavior indicates that for the polymer melt the
local viscous response is independent of chain length at short wave
lengths. 

\begin{center}
  \begin{figure}
    \scalebox{0.4}{
      \includegraphics{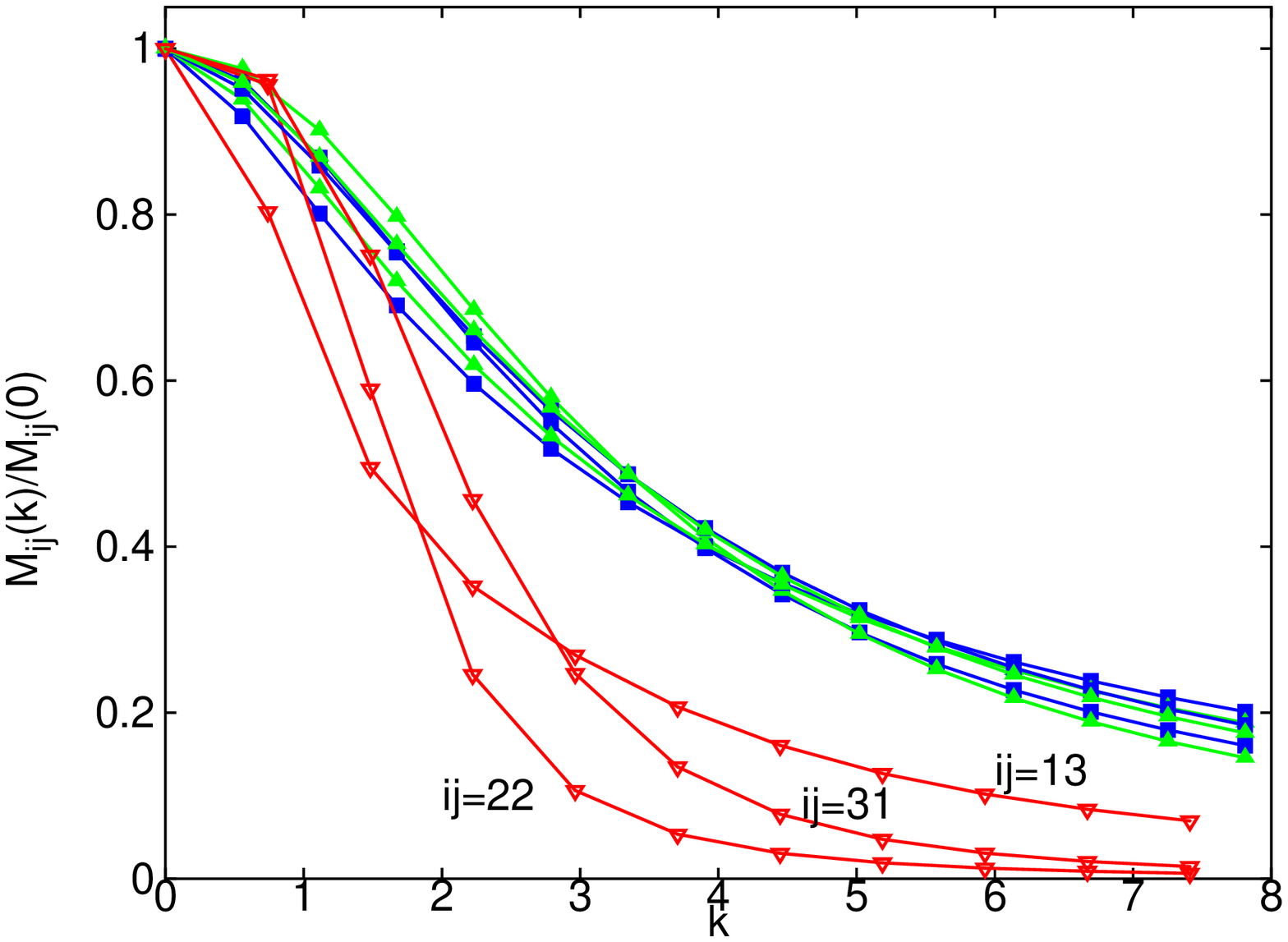}
    }
    \caption{
      \label{fig:4}
      McAllister coefficients as function of wave vector, $k$. Filled
      squares represent the results for the Kob-Andersen mixture,
      upward pointing triangles the Lennard-Jones mixture and downward
      pointing triangles the polymer mixture. Lines serve as a guide to
      the eye.
    }
  \end{figure}
\end{center}

To discuss this special behavior further we have plotted the
McAllister coefficients as functions of wave vector in
Fig. \ref{fig:4}. 
We see that $M_{ij}(k)/M_{ij}(0),
\{ij\}=\{31\},\{22\}$ and $\{13\}$,   
fall on a master curve  in the case of KA and LJ mixtures. 
This means that the three 
functions, $M_{ij}(k)$, may be described by a single master function
which is directly proportional to any of the three $M_{ij}(k)$. 
For the polymer mixture, on the other hand, this is not the
case. Note, that all the McAllister coefficients follow a Lorentzian form,
see Eq. (\ref{eq:lorentz}), but for the polymer mixture the parameters
$\alpha$ and $\beta$ are dependent on the index $\{ij\}$.   

In this report we have investigated the multiscale hydrodynamical
viscous response as a function of fluid composition. This was done 
through the viscosity kernel that was computed via equilibrium 
molecular dynamics simulations. We studied three different mixtures,
namely, (i)a Kob-Andersen mixture, (ii) a Lennard-Jones mixture using 
Lorentz-Berthelot mixing rule, and (iii) a polymer melt mixture. 
We observed that the viscosity kernel is independent of the wave vector
for large wave vectors in the case of a polymer melt, that is, 
the hydrodynamical response at these length scales is independent of
the composition. This was not the case for the simple Kob-Andersen and
Lennard-Jones mixtures. The deviation from ideal mixing is low in the
case of the  Lennard-Jones mixture, i.e. the Lorentz-Berthelot mixing
rule  agrees reasonably well with the predictions from ideal mixing for
all wave vectors studied here.  This is not the case for Kob-Andersen
and polymer mixtures. Finally, the relative deviation from ideal
mixing is relatively wave vector independent in the case of the Kob-Andersen and
Lennard-Jones mixtures. However, for the polymer mixture this deviation
shows a strong wave vector dependency, since the ideal mixing rule does
not predict the largely wave vector independent behavior of the
kernel, Fig 1 c).   

J.S. Hansen wishes to acknowedge Lundbeckfonden for supporting this
work as a part of Grant No. R49-A5634. The authors also wishes to
thank Prof. Peter J. Daivis for useful comments.

\end{document}